\documentclass[a4paper,11pt]{article}

\usepackage{amsmath}
\usepackage{bm}
\usepackage{cite}
\usepackage{color}
\usepackage[pdftex]{graphicx}
\usepackage{subfigure}
\usepackage{nicefrac}
\usepackage{multirow}
\usepackage{setspace}
\def \mb{\mathbf}                

\begin{document}

\title{\textbf{Optimal control-based inverse determination of electrode distribution for electroosmotic micromixer}}

\author{Yuan Ji$^{\dagger,\star}$, Yongbo Deng$^\dagger$\footnote{yongbo\_deng@hotmail.com}, Zhenyu Liu$^\ddagger$, Teng Zhou$^\S$, Yihui Wu$^\dagger$\\
$\dagger$ State Key Laboratory of Applied Optics \\
Changchun Institute of Optics, Fine Mechanics and Physics (CIOMP) \\
$\star$ University of Chinese Academy of Sciences, Beijing, China, 100039\\
$\S$ Mechanical and Electrical Engineering College, \\
Hainan University, Haikou 570228, China \\
$\ddagger$ Changchun Institute of Optics, Fine Mechanics and Physics (CIOMP) \\
Chinese Academy of Sciences, 130033, Changchun, Jilin, China}

\maketitle

\begin{abstract}
This paper presents an optimal control-based inverse method used to determine the distribution of the electrodes for the electroosmotic micromixers with external driven flow from the inlet. Based on the optimal control method, one Dirichlet boundary control problem is constructed to inversely find the optimal distribution of the electrodes on the sidewalls of electroosmotic micromixers and achieve the acceptable mixing performance. After solving the boundary control problem, the step-shaped distribution of the external electric potential imposed on the sidewalls can be obtained and the distribution of electrodes can be inversely determined according to the obtained external electric potential. Numerical results are also provided to demonstrate the effectivity of the proposed method.
\end{abstract}
\textbf{keyword}: Electroosmotic micromixer, electrode distribution, optimal control

\section{Introduction}\label{Introduction}
Lab on a chip is the generic term for the integration of the microdevices to carry out conventional analytical laboratory tests. Such devices offer significant benefits over traditional laboratory tests in terms of device size, sample/reagent usage, and can provide much faster results for chemical and biochemical analyses \cite{Dittrich2006,Manz2001}. Because of these advantages, lab on a chip devices are considered a promising option for the development of miniaturized devices for the environmental and defense monitoring, chemical synthesis and biomedical applications. In lab on a chip systems, various subcomponents such as pumps, mixers, reactors, and dilution chambers are integrated. Therefore, the study of fluid flow in microscale, i.e. microfluidics, has become central to the development of lab on a chip devices \cite{Stone2004,Squires2005,OuYang2014}. For lab on a chip devices, micromixers are often vital components as mixing is required for chemical applications, biological applications, and detection/analysis of chemical or biochemical content \cite{Jeong2010,Stroock2002,Li2007}.

Owing to small channel dimensions and low flow rates, the Reynolds number for flows in microfluidic device is typically very small. Hence, mixing through turbulent flow induced by inertial/viscous effects for aqueous solutions is not feasible in these miniaturized devices. Therefore, diffusion is the dominant mechanism in micromixing due to the absence of turbulence. Although it is difficult to induce turbulence in microchannels, an effective mixing in low Reynolds number flow regimes can be obtained by the chaotic advection mechanism, which can occur in regular smooth flows \cite{Ottino1989} and provides an effective increase in the interfacial contact area \cite{Deng2012}. Electroosmosis is one of the most common nonmechanical means of achieving chaotic advection in microfluidics. When a charged solid surface comes in contact with an electrolyte, an electric double layer (EDL) of ions is formed due to the interplay between electrical and diffusive forces \cite{Ammam2012}. The flow of liquids containing dissolved ions under the influence of electrical body forces is known as electroosmosis; it is a subject treated in depth in the electrokinetic transport literature \cite{Hunter1981,Qian2012}. A simple method has been proposed for mixing low Reynolds number electroosmotic flows in microchannels with patterned grooves \cite{Johnson2002}. These grooves induce spiral circulations around the flow axis at low Reynolds numbers and stretch and fold the streams with the result that a complete mixing can be achieved within a short mixing length. The use of unstable electrokinetic flow to achieve chaotic mixing effect has also been presented in \cite{Oddy2001,Lin20041,Chen2005,Shin2005,Qian2002}. Several numerical analytical investigation on electroosmotic mixing have been performed  \cite{Chang2006,Zhang2006}. And the mixing efficiency has been enhanced based on the periodic electroosmotic flow \cite{Lim2010}, modulation of electric fields \cite{Pacheco2008}, and shape optimization \cite{Jain2009} et al. The pattern of electroosmotic flow is mainly determined by the electrode distribution. For the complexity of electroosmotic flow, physical intuition-based determination of electrode distribution has its limitation. To overcome this limitation, it is necessary to develop the inverse termination method for electrode distribution of electroosmotic mciromixer.

In the electroosmotic micromixer, the mixing efficiency is mainly determined by the electrode distribution used to carry the externally applied electric potential. Therefore, this paper is focused on the method used to inversely determine the electrode distribution for electroosmotic micromixers. The discussed inverse termination method is built based in the optimal control method, which has been utilized to implement airfoil design \cite{}, sensor deployment \cite{}, control the convection diffusion \cite{Yuecel2015} and electric field for electrorheological fluids \cite{Reyes2015}. Based on the optimal control method, one boundary control problem is constructed for the electroosmotic micromixer in this paper. After solving of the problem, the electrode distribution can be determined according to the obtained step-shaped externally applied electric potential.

\section{Methodology}\label{Theory}
\subsection{Modeling}\label{Modeling}
When a micromixer is used to mix two fluidic flows with different solutes, the desired effect is the mixing of the two flows with anticipated concentration distribution at the outlet of the micromixer. The anticipated concentration distribution at the outlet can be specified by the designer based on the desired performance of the micromixer. The mixing performance of
the micromixer can be measured by the least square variance between the obtained concentration $c$ and the anticipated concentration $c_a$ at the outlet, named mixing measurement \cite{Li2007,Danckwerts1952}
\begin{equation}\label{Obj}
\begin{split}
    \Psi\left( c \right) = \int_{\Gamma_o} \left(c - c_a\right)^2\,\mathrm{d}\Gamma \bigg/ \int_{\Gamma_i} \left(c_r - c_a\right)^2\,\mathrm{d}\Gamma
\end{split}
\end{equation}
where $\Gamma_i$ and $\Gamma_o$ are the inlet and outlet of the micromixer, respectively; $c_r$ is the reference concentration distribution, which is usually chosen to be the given concentration distribution at the inlet. For micromixers, the required performance is to achieve the sufficient mixing
of the two solutes. Therefore, the anticipated concentration $c_a$ is specified to be the ideal concentration distribution of the solute at the outlet after sufficient mixing. In a electroosmotic micromixer with fixed geometry, the mixing efficiency is determined by the distribution of the external electric potential induced by the electrode potential. And the distribution of the electrode potential lies on the distribution of the electrodes at the sidewalls of the electroosmotic micromixer. Then the problem is on how to find a reasonable distribution of the electrodes that minimizes the mixing measurement and achieves sufficient mixing in a electroosmotic micromixer. In this paper, the optimal control method is adopted and one Dirichelet boundary control problem is constructed to solve this problem.

\begin{figure}[!htbp]
  \centering
  \includegraphics[width=0.6\columnwidth]{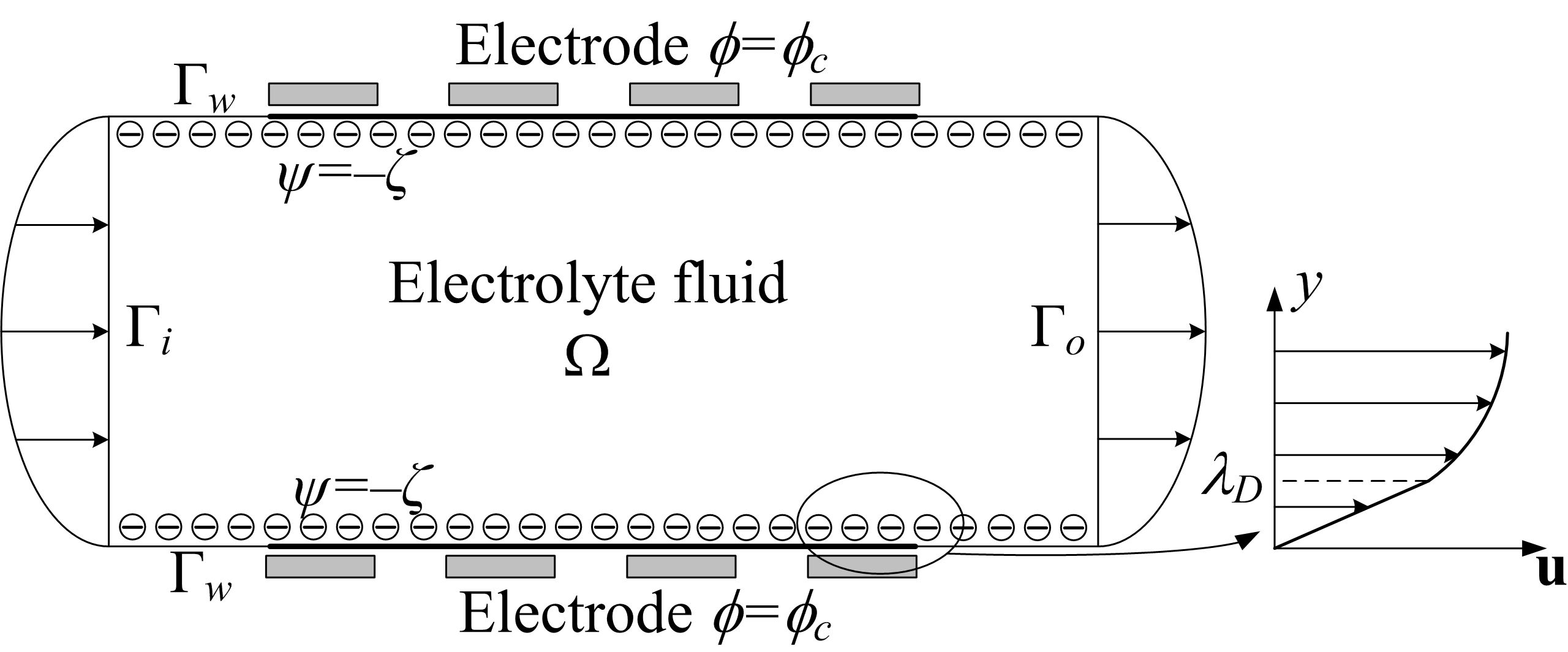}\\
  \caption{Schematic for the electroosmotic flow in the micromixer.}\label{SchematicEO}
\end{figure}

Under the precondition of the continuum assumption, the electroosmotic flow is described by the Navier-Stokes equations modified to include an electrical driving force term to represent the interaction between the excess ions of the electrical double layer (EDL) and the external electric field induced by the electrode potential, where an assumption is made that the Joule heating effect is negligible and can be ignored \cite{Probstein1994}. In electroosmotic flows, the electric potential can be decomposed into an external electric potential due to the imposition of the externally applied electrode potential and an electric potential due to surface wall charge \cite{Patankar1998}. Then the body force imposes on the fluid is the electric force of these two potentials. Based on the above description, the governing equations of the electroosmotic flow are
\begin{equation}\label{NSEqu}
\begin{split}
& \rho \mb u \cdot \nabla \mb u = \nabla \cdot \left[-p \mb I + \eta \left( \nabla \mb u + \nabla \mb u^\mathrm{T} \right) \right] + {\varepsilon \over {\lambda_D^2}} \psi \nabla \phi,~\mathrm{in}~\Omega \\
& -\nabla \cdot \mb u = 0,~\mathrm{in}~\Omega \\
\end{split}
\end{equation}
where $\mb u$ is the fluid velocity; $p$ is the fluid pressure; $\rho$ and $\eta$ are the density and viscosity of the fluid, respectively; $\lambda_D$ is the Debye length, which is the characteristic thickness of the EDL for a given solid-electrolyte liquid interface; $\varepsilon$ is the dielectric constant of the electrolyte solution; $\psi$ is the electric potential due to surface wall charge; $\phi$ is the external electric potential; $\Omega$ is the space domain occupied by the electroosmotic flow, and the boundaries of $\Omega$ include the inlet port $\Gamma_i$, the outlet port $\Gamma_o$ and the sidewalls $\Gamma_w$ (Fig. \ref{SchematicEO}). The imposed boundary conditions for the Navier-Stokes equations are
\begin{equation}\label{FlowBnd}
\begin{split}
    & \mb u = \mb u_i,~\mathrm{on}~ \Gamma_{i} \\
    & \mb u = \mb 0,~\mathrm{on}~\Gamma_{w} \\
    & \left[-p \mb I + \eta \left( \nabla \mb u + \nabla \mb u^\mathrm{T} \right) \right] \cdot \mb n =  \mb 0,~\mathrm{on}~\Gamma_{o} \\
\end{split}
\end{equation}
where $\mb u_i$ is a given velocity distribution at the inlet port; $\mb n$ is the unit outward normal vector on the boundary of $\Omega$.
In micromixing, the two factors that influence the mixing performance of a micromixer are diffusion and chaotic advection. And the mixing of flows is described using the convection-diffusion equation
\begin{equation}\label{CDEqu}
\begin{split}
\mb u \cdot \nabla c = D \nabla^2 c,~\mathrm{in}~\Omega \\
\end{split}
\end{equation}
where $D$ is the diffusion constant of the fluid. The imposed boundary conditions for the convection-diffusion equation are
\begin{equation}\label{CDEBnd}
\begin{split}
    & c = c_i\left(\mb x\right),~\mathrm{on} ~ \Gamma_{i} \\
    & \nabla c \cdot \mb n = 0,~\mathrm{on}~ \Gamma_w \cup \Gamma_o \\
\end{split}
\end{equation}
where $c_i$ is the given concentration distribution at the inlet port of the electroosmotic micromixer.
For a symmetrical and univalent electrolyte at room temperature, the Debye length is on the magnitude $10\mathrm{nm}$ for a concentration of $10^{-3}\mathrm{M}$. In other words, the Debye length is very small compared to the characteristic length of the microchannel. Moreover, within the EDL, the electrical potential drops from the zeta potential to zero \cite{Probstein1994,Hunter1981}. In general, the zeta potential is of the order of $0.1\mathrm{V}$. The ion distribution in the EDL is influenced primarily by the zeta potential, and the distribution of the potential due to surface wall charge can be obtained by solving the equation
\begin{equation}\label{WallChargePotential}
\begin{split}
\nabla^2 \psi = {1 \over {\lambda_D^2}} \psi,~\mathrm{in}~\Omega
\end{split}
\end{equation}
For equation \ref{WallChargePotential}, the imposed boundary conditions are
\begin{equation}\label{ZetaBnd}
\begin{split}
    & \psi = -\zeta,~\mathrm{on} ~ \Gamma_w \\
    & \nabla \psi \cdot \mb n =  0,~\mathrm{on}~ \Gamma_i \cup \Gamma_o \\
\end{split}
\end{equation}
where $\zeta$ is the zeta potential. Since the external electric potential arises from external charges, it satisfies Laplacian equation within the fluid domain
\begin{equation}\label{Laplacian}
\begin{split}
\nabla^2 \phi = 0,~\mathrm{in}~\Omega
\end{split}
\end{equation}
and the corresponding boundary conditions are
\begin{equation}\label{LaplacianBnd}
\begin{split}
    & \phi = \phi_c\left(\mb x\right),~\mathrm{on} ~ \Gamma_{w} \\
    & \nabla \phi \cdot \mb n =  0,~\mathrm{on}~ \Gamma_i \cup \Gamma_o \\
\end{split}
\end{equation}
where $\phi_c$ is the electrode potential on the sidewalls of the electroosmotic micromixer. Then the micromixing in the electroosmotic flow can be described using the coupled system of equation \ref{NSEqu}, \ref{CDEqu}, \ref{WallChargePotential}, and \ref{Laplacian}.

\begin{figure}[!htbp]
  \centering
  \includegraphics[width=0.6\columnwidth]{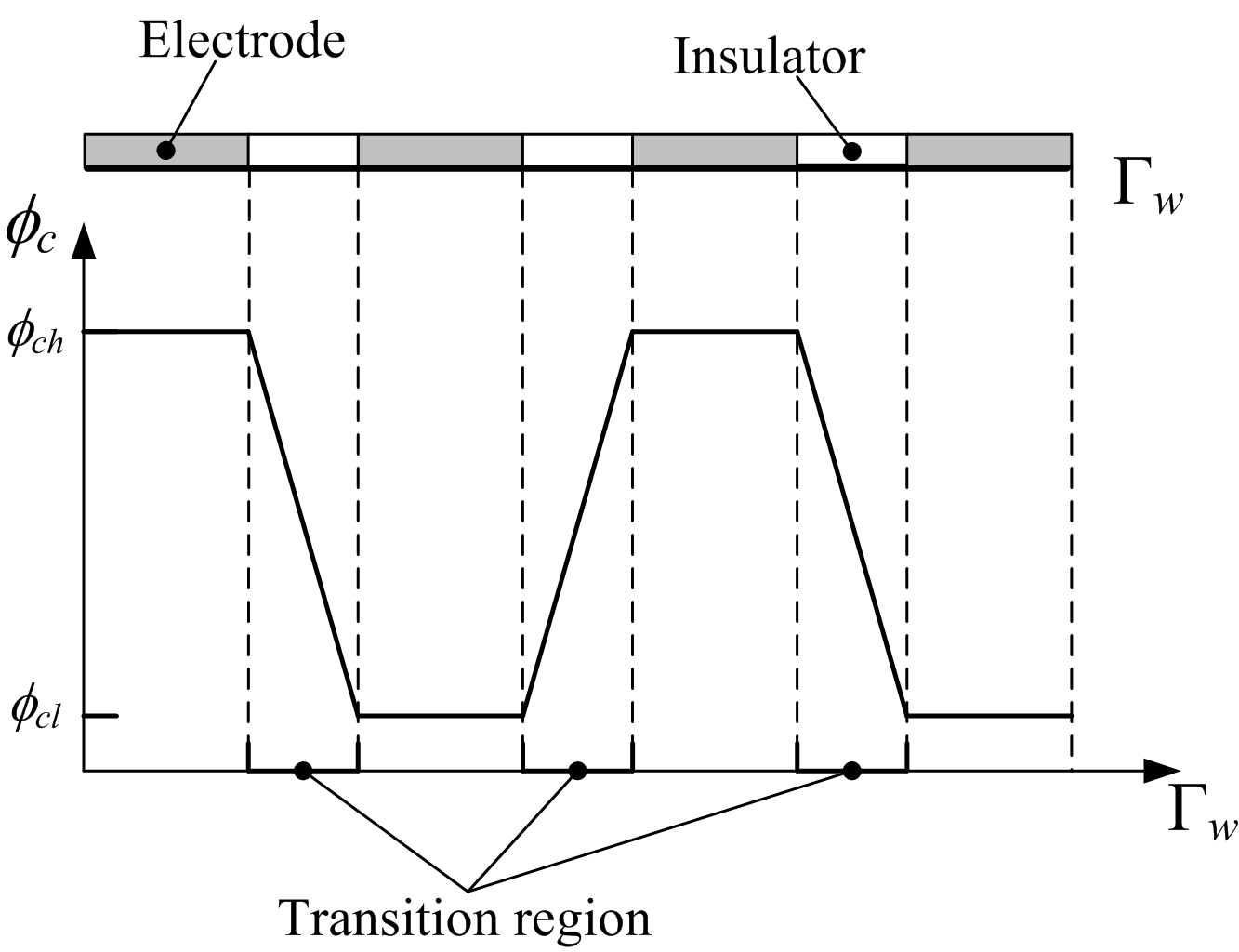}\\
  \caption{Schematic for the electrode and the corresponding external electric potential at the sidewall of the electroosmotic micromixer.}\label{SchematicElectrode}
\end{figure}

Based on the above description, the optimal control problem, used to find the reasonable distribution of the electrode potential and minimize the mixing measurement, can be constructed with the mixing measurement as objective, the coupling system of equation \ref{NSEqu}, \ref{CDEqu}, \ref{WallChargePotential} and \ref{Laplacian} as constraints, and the electrode potential as control variable.
Because the control variable (i.e. electrode potential) is defined on the sidewalls which is the Dirichlet boundary of the coupled system, the constructed optimal control problem is a Dirichlet boundary control problem. In the optimal control problem, the admittable set of the control variable is set to be $\left[\phi_{cl},\phi_{ch}\right]$, where the values of $\phi_{cl}$ and $\phi_{ch}$ can be determined due to the engineering reality.
In order to ensure the manufacturability of the obtained electrode distribution, the distribution of the electrode potential corresponding to the electrode distribution should satisfy the conditions as demonstrated in Fig. \ref{SchematicElectrode}: the electrode potential on every electrode should be an electric level corresponding to the constant potential $\phi_{cl}$ or $\phi_{ch}$; the size of the transition region, filled with insulators, between two neighboring electrodes, should be large enough to avoid excess high electric field strength and capacitor breakdown. These conditions can be ensured using the filter and projection methods and imposing a constraint on the electric field strength, where the control variable is filtered using the Helmholtz filter and the filtered variable is projected using the threshold method in this paper \cite{Lazarov2010,Guest2004,Deng2014}. The control variable is evolved using the robust numerical optimization algorithm MMA (the method of moving asymptotes) \cite{Svanberg1987,Borrvall2003}. Based on the filtering of the control variable, the reasonable distance between two neighboring electrodes at the sidewall can be ensured, and the filter is implemented by solving the following Helmholtz type PDE
\begin{equation}\label{HelmFilter}
\begin{split}
    & -r^2\nabla_{\Gamma}^2 \widetilde{\phi}_c + \widetilde{\phi}_c = \phi_c,~\mathrm{on}~\Gamma_w \\
    & \mb n_{\Gamma} \cdot \nabla_{\Gamma}\widetilde{\phi}_c = 0,~\mathrm{at}~\partial\Gamma_w
\end{split}
\end{equation}
where $\widetilde{\phi}_c$ is the filtered control variable; $r$ is the filter radius; $\nabla_\Gamma$ is the gradient operator defined on $\Gamma_w$; $\mb n_\Gamma$ is the unit outward normal vector on the boundary $\Gamma_w$. The distance between two neighboring electrodes can be controlled by reasonably choosing the value of the filter radius to control the size of the transition region. Generally, higher value of filter radius corresponds to larger size of the transition region. The threshold projection can ensure the change of the external electric potential as linear as possible in the transition region between two neighboring electrodes on the sidewall, and it is performed using the following formulation:
\begin{equation}\label{ThresholdProjection}
\begin{split}
    \overline{\widetilde{\phi}}_c = {{\mathrm{tanh}\left(\beta\xi\right) + \mathrm{tanh}\left(\beta\left(\widetilde{\phi}_c - \xi\right)\right)}\over{\mathrm{tanh}\left(\beta\xi\right) + \mathrm{tanh}\left(\beta\left(1 - \xi\right)\right)}}
\end{split}
\end{equation}
where $\overline{\widetilde{\phi}}_c$ is the projected control variable; $\xi\in\left[0,1\right]$ and $\beta$ are the threshold and projection parameters for the threshold projection, respectively. On the choice of the values of $\xi$ and $\beta$, one can refer to \cite{Wang2011}. Using the threshold projection, the filtered control variable can also be projected to $\phi_{cl}$ or $\phi_{ch}$ at the points in the region corresponding to the electrodes, and the interim values in $\left(\phi_{cl},\phi_{ch}\right)$ are avoided effectively, i.e. the external electric potential applied on the control boundary will only have the constant values $\phi_{cl}$ and $\phi_{ch}$, which can be realized by fabricate separated electrodes on the sidewall of the electroosmotic micromixer. To avoid the excess high electric field strength, the electric field strength induced by the external electric potential is constrained as
\begin{equation}\label{ConstraintE}
    \int_\Omega \left|\nabla \phi\right|^2\,\mathrm{d}\Omega \leq C_0
\end{equation}
where $C_0$ is a constant, chosen based on numerical experiments and engineering reality.
Then, the manufacturability of the design corresponding to the result of the optimal control problem is ensured based on the Helmholtz filter, threshold projection and electric field strength constraint. For summary, the optimal control problem for inverse determination of the electrode distribution for electroosmotic micromixer can be constructed to be
\begin{equation}\label{OptContrPro}
\begin{split}
& \mathrm{min}~\Psi\left( c \right) = \int_{\Gamma_o} \left(c - c_a\right)^2\,\mathrm{d}\Gamma \bigg/ \int_{\Gamma_i} \left(c_r - c_a\right)^2\,\mathrm{d}\Gamma \\
& \mathrm{s.t.} \left\{
\begin{split}
& \rho \mb u \cdot \nabla \mb u = \nabla \cdot \left[-p \mb I + \eta \left( \nabla \mb u + \nabla \mb u^\mathrm{T} \right) \right] + {\varepsilon \over {\lambda_D^2}} \psi \nabla \phi,~\mathrm{in}~\Omega \\
& -\nabla \cdot \mb u = 0,~\mathrm{in}~\Omega \\
& \mb u \cdot \nabla c = D \nabla^2 c,~\mathrm{in}~\Omega \\
& \nabla^2 \psi = {1 \over {\lambda_D^2}} \psi,~\mathrm{in}~\Omega\\
& \int_\Omega \left|\nabla \phi\right|^2\,\mathrm{d}\Omega \leq C_0
\end{split}\right.
\end{split}
\end{equation}
By solving the optimal control problem, the electrode distribution corresponding to the external electric potential can be determined and minimal of the mixing measurement can be derived.

\subsection{Analyzing and solving}\label{Analyzing}
The constructed optimal control problem in section \ref{Modeling} is an optimization problem with partial differential equation constraints, and it can be analyzed using the adjoint method \cite{Hinze2009}. In this paper, the optimal control problem is solved by the finite element method. To use
the linear elements for the partial differential equations, the Navier-Stokes equations and convection-diffusion equation are stabilized using the generalized least squares (GLS) and the streamline upwind Petrov-Galerkin (SUPG) technologies, respectively \cite{Donea2003}. Then the stabilized weak forms are
\begin{equation}\label{WeakNS}
\begin{split}
        & \int_\Omega \rho \mb u \cdot \nabla \mb u \cdot \mb v + \int_\Omega \left[-p \mb I + \eta \left( \nabla \mb u + \nabla \mb u^\mathrm{T} \right) \right] : \nabla \mb v \,\mathrm{d}\Omega - \int_\Omega {\varepsilon \over {\lambda_D^2}} \psi \nabla \phi \cdot \mb v \,\mathrm{d}\Omega - \int_\Omega q \nabla \cdot \mb u \,\mathrm{d}\Omega \\
        & + \sum_{i=1}^{N_e} \int_{\Omega_i} \tau_{GLS} \nabla p \cdot \nabla q \,\mathrm{d}\Omega = 0,
        ~\forall \mb v \in \mathbf{H}_1\left(\Omega\right),~\forall q \in \mathrm{L}_2\left(\Omega\right)\\
        & \mb u = \mb u_i, ~\mathrm{on}~\Gamma_i \\
        & \mb u = \mb 0,~\mathrm{on}~\Gamma_w
\end{split}
\end{equation}
for the Navier-Stokes equations, and
\begin{equation}\label{WeakCD}
\begin{split}
        & \int_\Omega \mb u \cdot \nabla c~s \,\mathrm{d}\Omega + \int_\Omega D \nabla c \cdot \nabla s \,\mathrm{d}\Omega + \sum_{i=1}^{N_e} \int_{\Omega_i} \tau_{SUPG} \mb u \cdot \nabla s \left(\mb u \cdot \nabla c - D\nabla^2 c \right) \,\mathrm{d}\Omega = 0, \\
        & \forall s \in \mathrm{H}_1\left(\Omega\right)\\
        & c = c_i,~\mathrm{on}~\Gamma_i
\end{split}
\end{equation}
for the convection-diffusion equation,
where $\mathrm{H}_1\left(\Omega\right)$ and $\mathrm{L}_2\left(\Omega\right)$ are the first order Sobolev space and the second-order Lebesgue integrable functional space, respectively; $N_e$ is the number of elements used to discretize the computational domain; and $\Omega_i$ is the domain of the i-th element; $\tau_{GLS}$ and $\tau_{SUPG}$ are the stabilization parameters. The stabilization parameters are chosen according to \cite{Donea2003,Andreasen2009}
\begin{equation}\label{Para}
\begin{split}
& \tau_{GLS} = {{h^2}\over{12\eta}}\\
& \tau_{SUPG} = \left({4\over{h^2D}}+{{2\left|\mb u\right|}\over{h}}\right)^{-1}\\
\end{split}
\end{equation}
where $h$ is the element size.
Based on the adjoint analysis of the objective in equation \ref{Obj}, the weak form of the adjoint equations of the convection-diffusion equation, the Navier-Stokes equations and the Laplacian equation are obtained as: find $c_a\in\mathrm{H}_1\left(\Omega\right)$,
 $\mb u_a\in\mathbf{H}_1\left(\Omega\right)$, $p_a\in\mathrm{L}_2\left(\Omega\right)$ and $\phi_a\in\mathrm{H}_1\left(\Omega\right)$ satisfying
\begin{equation}\label{AdjCD}
\begin{split}
        & \int_\Omega \left(\mb u \cdot \nabla s~c_a + \nabla s \cdot \nabla c_a \right) \,\mathrm{d}\Omega + \sum_{i=1}^{N_e} \int_{\Omega_i} \tau_{SUPG} \mb u \cdot \nabla c_a \left( \mb u \cdot \nabla s - D\nabla^2 s \right) \,\mathrm{d}\Omega \\
        & + \int_{\Gamma_o} 2 \left(c - c_a\right)s\,\mathrm{d}\Gamma \bigg/ \int_{\Gamma_i} \left(c_r - c_a\right)^2\,\mathrm{d}\Gamma = 0,~\forall s \in \mathrm{H}_1\left(\Omega\right) \\
        & c_a = 0,~\mathrm{on}~\Gamma_i
\end{split}
\end{equation}
for the convection-diffusion equation, and
\begin{equation}\label{AdjNS}
\begin{split}
      & \int_\Omega \big\{ \rho \left(\mb v \cdot \nabla \mb u + \mb u \cdot \nabla \mb v \right) \cdot \mb u_a + \left[ \eta \left( \nabla \mb v + \nabla \mb v^\mathrm{T} \right) - q \mb I\right] : \nabla \mb u_a - p_a \nabla \cdot \mb v \big\} \,\mathrm{d}\Omega + \\
      & \sum_{i=1}^{N_e} \int_{\Omega_i} \tau_{GLS} \nabla q \cdot \nabla p_a \,\mathrm{d}\Omega = - \int_\Omega \mb v \cdot \nabla c~c_a \,\mathrm{d}\Omega - \\
      & \sum_{i=1}^{N_e} \int_{\Omega_i} \bigg[\left({{\partial \tau_{SUPG}}\over{\partial \mb u}} \cdot \mb v \right) \left(\mb u \cdot \nabla c_a \right) \left(\mb u \cdot \nabla c - D\nabla^2 c \right) \\
      & + \tau_{SUPG} \mb v \cdot \nabla c_a \left(\mb u \cdot \nabla c - D\nabla^2 c \right) + \tau_{SUPG} \left(\mb u \cdot \nabla c_a\right) \left( \mb v \cdot \nabla c \right) \bigg]\,\mathrm{d}\Omega,~\forall \mb v \in \mathbf{H}_1\left(\Omega\right),\\
      & \forall q \in \mathrm{L}_2\left(\Omega\right)\\
      & \mb u_a = \mb 0,~\mathrm{on}~\Gamma_i\cup\Gamma_w
\end{split}
\end{equation}
for the Navier-Stokes equations, and
\begin{equation}\label{AdjZeta}
\begin{split}
&-\int_\Omega \nabla \phi_a \cdot \nabla \varphi \,\mathrm{d}\Omega + \int_\Omega {\varepsilon \over {\lambda_D^2}} \nabla\cdot \left(\psi\mb u_a\right) \varphi \,\mathrm{d}\Omega -
\int_{\Gamma_o} {\varepsilon \over {\lambda_D^2}} \psi \mb u_a \cdot \mb n \varphi \,\mathrm{d}\Gamma = 0, ~ \forall \varphi \in \mathrm{H}_1
\left(\Omega\right)\\
& \phi_a = 0,~\mathrm{on}~\Gamma_w\\
\end{split}
\end{equation}
for the Laplacian equation, where $c_a$, $\mb u_a$, $p_a$ and $\phi_a$ are the adjoint variables corresponding to $c$, $\mb u$, $p$ and $\phi$, respectively. In the adjoint analysis, equation \ref{WallChargePotential} need not to be included, because the potential due to surface wall charge is independent of the externally applied potential. The adjoint sensitivity of the optimal control problem can be obtained as
\begin{equation}\label{AdjDrv}
\begin{split}
   \delta \hat \Psi = \int_{\Gamma_w} - \nabla\phi_a\cdot\mb n \left(\phi_{ch}-\phi_{cl}\right) {{\mathrm{d}\overline{\widetilde{\phi}}_c}\over{\mathrm{d}\widetilde{\phi}_c}}
   {{\mathrm{d}\widetilde{\phi}_c}\over{\mathrm{d}\phi_c}}
   \delta\phi_c\,\mathrm{d}\Gamma
\end{split}
\end{equation}
For the constraint in equation \ref{ConstraintE}, the weak form of the adjoint equation is
\begin{equation}\label{AdjEquCst}
\begin{split}
& \int_\Omega \nabla\left(\phi_a-\phi\right)\cdot\nabla\psi\,\mathrm{d}\Omega = 0,~\forall \varphi \in \mathrm{H}_1\left(\Omega\right) \\
& \phi_a = 0,~\mathrm{on}~\Gamma_w
\end{split}
\end{equation}
and the adjoint sensitivity is
\begin{equation}\label{AdjSensCst}
    \delta C = \int_{\Gamma_w} \nabla \left(\phi-\phi_a\right) \cdot \mathbf{n} \left(\phi_{ch}-\phi_{cl}\right) {{\mathrm{d}\overline{\widetilde{\phi}}_c}\over{\mathrm{d}\widetilde{\phi}_c}}
   {{\mathrm{d}\widetilde{\phi}_c}\over{\mathrm{d}\phi_c}}
   \delta\phi_c\,\mathrm{d}\Gamma
\end{equation}
In the discretization of the sensitivities in equation \ref{AdjDrv} and \ref{AdjSensCst}, ${{\mathrm{d}\widetilde{\phi}_c}\over{\mathrm{d}\phi_c}}$ should be treated skillfully to avoid the inverse of matrix, for details one can refer to \cite{Lazarov2010}.

Solving of the optimal control problem is implemented using the gradient-based iterative approach. In the iterative procedure, the coupled system of equation \ref{NSEqu}, \ref{CDEqu}, \ref{WallChargePotential} and \ref{Laplacian}, and the corresponding adjoint equations in the weak
form are solved by the finite element method using the commercial software COMSOL Multiphysics (version 3.5) with linear elements (http://www.comsol.com). Then, the adjoint derivative can be obtained according to equation \ref{AdjDrv}. The discretized control variable is updated using MMA until the convergence criteria is satisfied, where the convergence criteria is set to be the maximal change of the control variable in consecutive $5$ iterations less than $1\times10^{-3}$ or the maximal iteration number $400$.
\begin{table}[htbp]
\centering
\begin{tabular}{l}
  \hline
  1. Give the initial value of the control variable $\phi_c$; \\
  2. Solve the coupled system of equation \ref{NSEqu}, \ref{CDEqu}, \ref{WallChargePotential}, and \ref{Laplacian} by the finite element method;\\
  3. Solve the weak form adjoint equations (equation \ref{AdjCD}, \ref{AdjNS}, \ref{AdjZeta} and \ref{AdjEquCst}); \\
  4. Compute the adjoint derivatives (equation \ref{AdjDrv} and \ref{AdjSensCst}) and \\
  ~~~~the corresponding objective and constraint values;\\
  5. Update the control variable by MMA; \\
  6. Check for convergence; if the stopping conditions are not satisfied, go to 2; and\\
  7. Post processing\\
  \hline
\end{tabular}
\caption{Procedure of the iterative approach for solving the optimal control problem.}
\label{FlowchartMixer}
\end{table}

\section{Results and discussion}\label{Results}
\begin{figure}[!htbp]
  \centering
  \includegraphics[width=0.8\columnwidth]{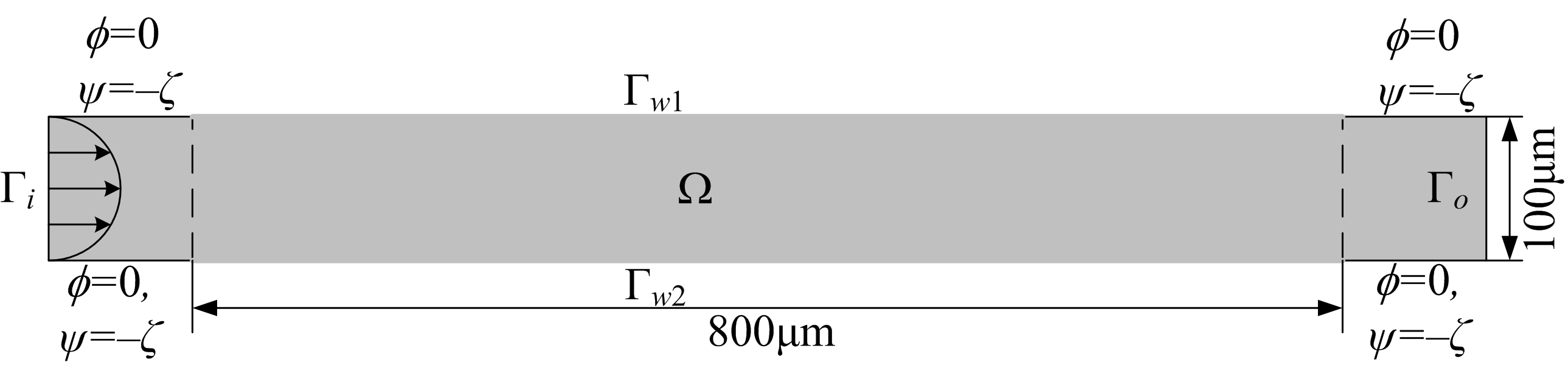}\\
  \caption{Schematic of the electroosmotic micromixer in a straight microchannel.}\label{2DSchematic}
\end{figure}

To demonstrate the effectivity of the proposed method used to inversly determine the electrode distribution for electroosmotic micromixers, the electroosmotic micromixer in a straight microchannel with externally electric potential imposed on the sidewalls is investigated numerically in the following. The schematic of the electroosmotic micromixer is shown in Fig. \ref{2DSchematic}, where the parabolic fluid velocity is loaded at the inlet $\Gamma_i$. The Reynolds number and P\'{e}clet number of the flow in the micromixer are $1$ and $1000$, respectively. The dielectric constant of the electrolyte solution, the Debye length and the zeta potential are set to be $7.4\times10^{-11} \mathrm{C^2/\left(N \cdot m\right)^2}$, $765\mathrm{nm}$ and $0.1\mathrm{V}$, respectively. The bounds of the external electric potential are set as $\phi_{cl} = 0\mathrm{V}$ and $\phi_{ch} = 200\mathrm{V}$. The upper bound of the constraint in equation \ref{ConstraintE} is chosen to be $C_0 = 5.7\times10^5$. Such choice of the parameter $C_0$ is to enforce the externally applied electric field strength no more than the general value $10^7\mathrm{V}/\mathrm{m}$ \cite{Chang2006}.

\begin{figure}[!htbp]
  \centering
  \subfigure[Electrode potential]
  {\includegraphics[width=0.8\columnwidth]{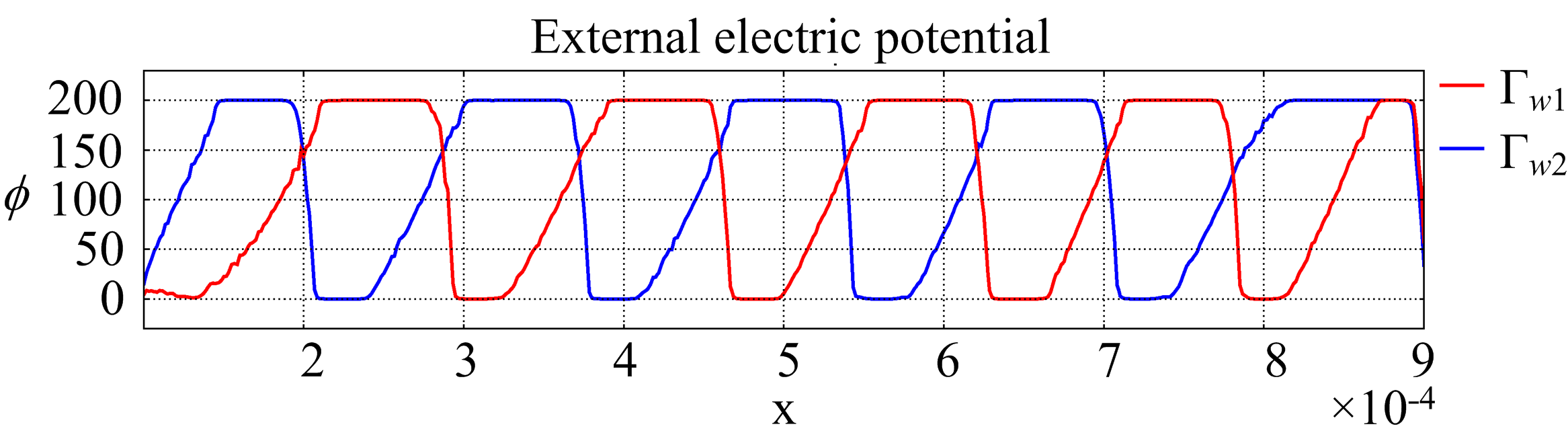}}\\
  \subfigure[Electrode distribution]
  {\includegraphics[width=0.8\columnwidth]{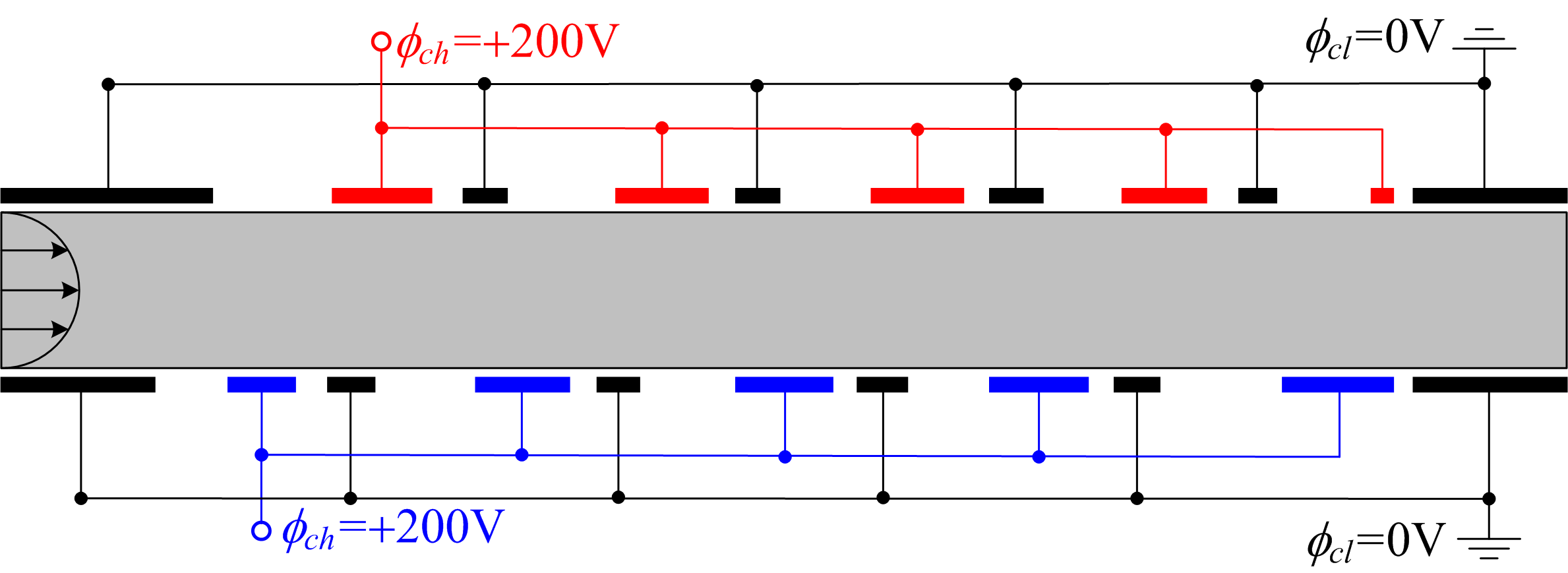}}\\
  \subfigure[Electric potential distribution]
  {\includegraphics[width=0.8\columnwidth]{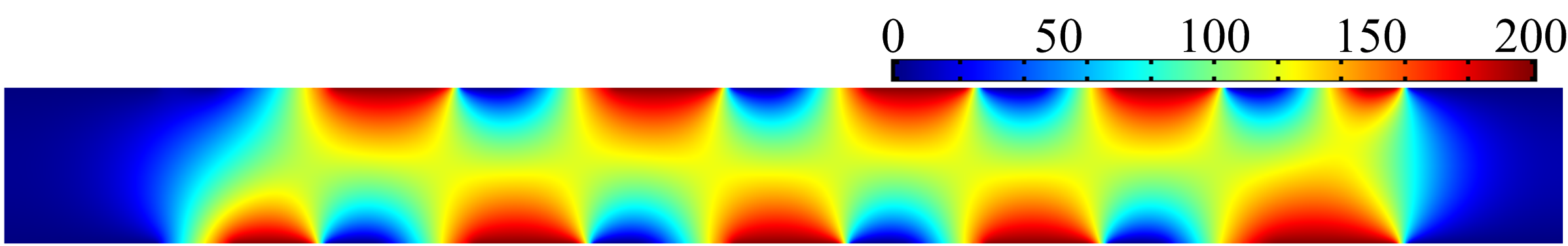}}\\
  \subfigure[Streamline distribution]
  {\includegraphics[width=0.8\columnwidth]{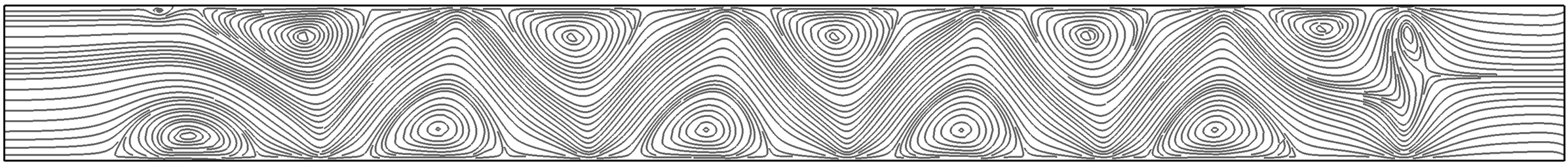}}\\
  \subfigure[Velocity distribution (red arrows) and anticipated concentration contour (blue curve)]
  {\includegraphics[width=0.8\columnwidth]{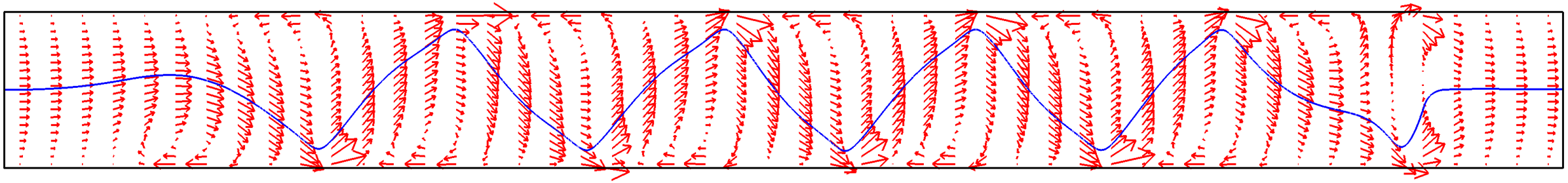}}\\
  \caption{(a) Electrode potential obtained using the optimal control method; (b) electrode distribution corresponding to the obtained electrode potential; (c) electric potential distribution induced by the obtained wall potential; (d) streamline distribution in the electroosmotic micromixer; (e) velocity distribution (red arrows) and anticipated concentration contour (blue curve) in the electroosmotic flow, where the chaotic advection and deformation of the interface between fluids is demonstrated.}\label{ElectricPotential}
\end{figure}

\begin{figure}[!htbp]
  \centering
  \subfigure[Concentration distribution without electrode potential]
  {\includegraphics[width=0.8\columnwidth]{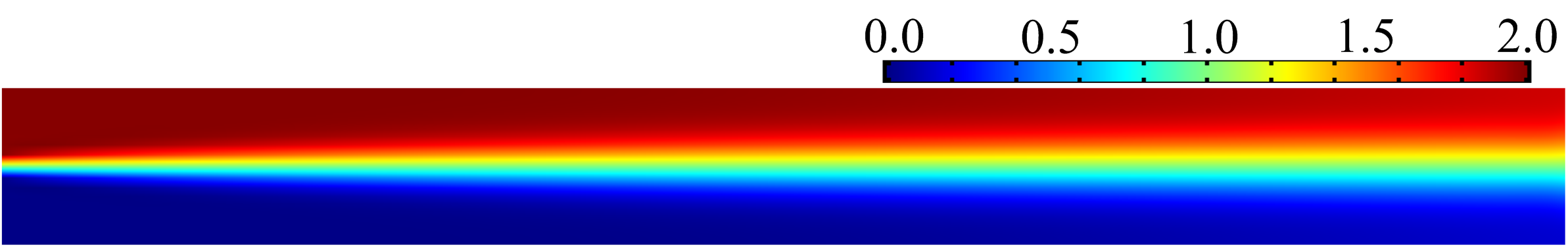}}\\
  \subfigure[Concentration distribution with the electrode potential shown in Fig. \ref{ElectricPotential}a]
  {\includegraphics[width=0.8\columnwidth]{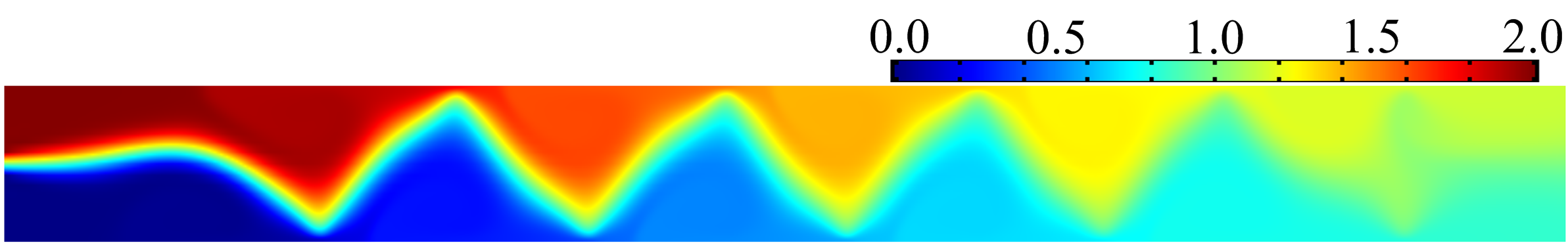}}\\
  \caption{(a) Concentration distribution in the microchannel without electrode potential, where the value of the mixing measurement is $0.6$; (b) concentration distribution with the electrode potential shown in Fig. \ref{ElectricPotential}a, and the corresponding value of the mixing measurement is 0.015, which is lower than the threshold level of mixing defined as 0.050 \cite{Li2007}. Therefore, complete mixing is achieved, when the electrode potential obtained using optimal control method is imposed on the sidewalls of the electroosmotic micromixer in Fig. \ref{2DSchematic}.}\label{Comparison}
\end{figure}

Based on the optimal control theory in Section \ref{Theory}, the optimal distribution of the electrode potential is obtained as shown in Fig. \ref{ElectricPotential}a. In Fig. \ref{ElectricPotential}a, the low and high levels correspond to the electrodes with electric potentials equal to $0\mathrm{V}$ and $200\mathrm{V}$, respectively; the declining parts between the neighboring low and high levels correspond to the regions filled with insulators used to separate neighboring electrodes. Therefore, the electrode distribution at the sidewalls of the electroosmotic micromixer can be determined according to the above analysis of the obtained optimal distribution of the externally applied electric potential (Fig. \ref{ElectricPotential}b). Fig. \ref{ElectricPotential}b shows that the electrodes has a interlaced arrangement. The interlaced arrangement of the low and high levels can avoid the counteraction of the electric force loaded on the electrolyte effectively. The distribution of the electric potential, induced by the electrode potential, is shown in Fig. \ref{ElectricPotential}c. From Fig. \ref{ElectricPotential}c, one can see that high gradient of electric potential (electric strength) is produced near the region between neighboring electrodes. The high electric potential gradient results in the large electric force load on the electrolyte. Therefore, the streamlines of the microflow is distorted impetuously, and vortexes arise along with the distortion of the streamlines in the straight microchannel (Fig. \ref{ElectricPotential}d). The distortion of the streamline and induced vortexes along the flow direction give rise to the enhancement of the chaotic advection, which is an interplay between the inertial, centrifugal, and viscous effects of the fluid flow. The enhancement of the chaotic advection deformed the interface between fluids strongly; the area of the interface grows exponentially; and diffusion becomes efficient (Fig. \ref{ElectricPotential}e). Therefore, the electrode distribution corresponding to the obtained electrode potential improves the micromixing effectively, and this can be confirmed based on the comparison between Fig. \ref{Comparison}a and \ref{Comparison}b.

\begin{figure}[!htbp]
  \centering
  \subfigure[Electric potential distribution]
  {\includegraphics[width=0.8\columnwidth]{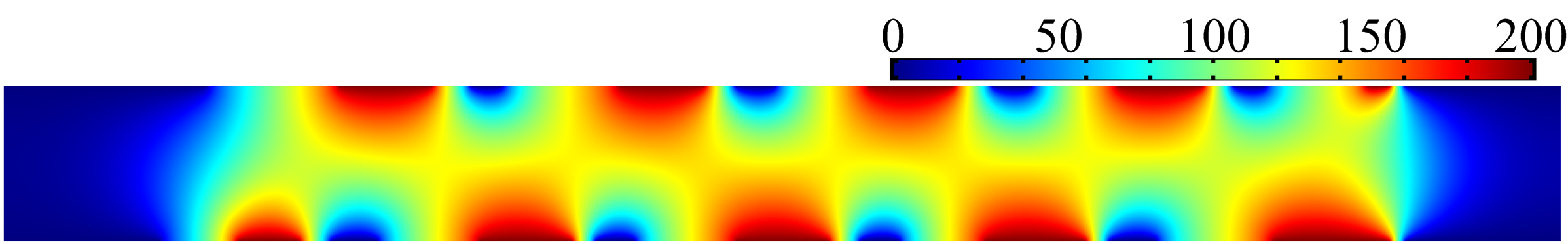}}\\
  \subfigure[Streamline distribution]
  {\includegraphics[width=0.8\columnwidth]{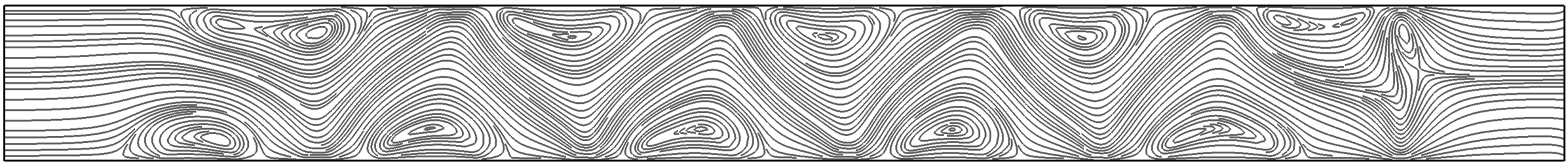}}\\
  \subfigure[Concentration distribution]
  {\includegraphics[width=0.8\columnwidth]{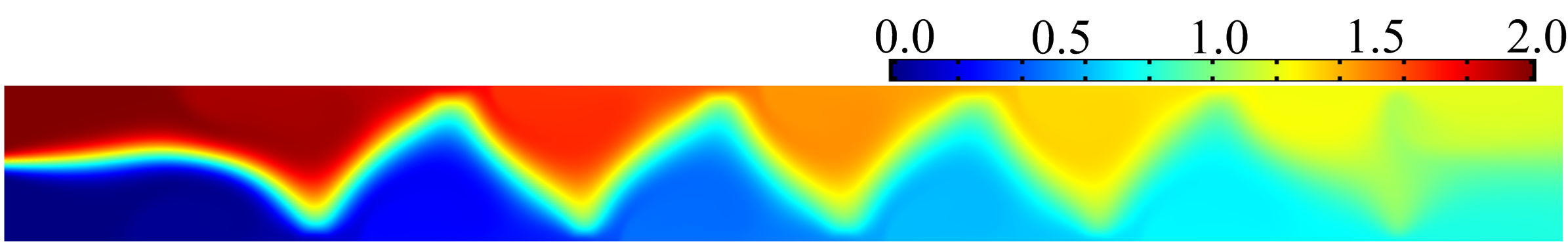}}\\
  \caption{(a) Electric potential distribution corresponding to the electrode distribution shown in Fig. \ref{ElectricPotential}b; (b) streamline distribution induced by the electrode distribution shown in Fig. \ref{ElectricPotential}b; (c) concentration distribution in the electroosmotic micromixer with electrode distribution as shown in Fig. \ref{ElectricPotential}b, and the value of the mixing measurement is $0.025$ lower than the mixing threshold $0.050$.}\label{PostProcessing}
\end{figure}

In the following, the postprocessing of the numerical results is performed. With the electrode distribution shown in Fig. \ref{ElectricPotential}b, the distribution of the electric potential, streamline and concentration are computed and shown in Fig. \ref{PostProcessing}. From the comparison between the results in Fig. \ref{ElectricPotential} and \ref{PostProcessing}, the consistency between the electric potential distributions corresponding to the optimal control method and the electrode distribution determined according to the electrode potential can be confirmed; and the effectivity of the proposed method used to determine the electrode distribution for electroosmotic micromixers is demonstrated furthermore.


\section{Conclusions}
In this paper, the inverse method used to determine the electrode distribution for electroosmotic micromixers has been proposed based on the optimal control method. The electrode distribution is inversely determined based on solving one optimal control problem to minimize the mixing measurement. And the optimal control problem is constrained by the governing equations of the electroosmotic micromixing. The control variable is set to be the electrode potential distribution applied on the sidewall of the electroosmotic micromixer. The electric field strength in the micromixer has also been constrained to avoid the capacitor breakdown phenomenon. Based on the adjoint analysis of the optimal control problem, the control variable is evolved using MMA to derive potential distribution with low and high levels, which correspond to the electrode on the sidewall of the electroosmotic micromixer. The manufacturability of the obtained electrode distribution is ensured by the filtering and projection of the control variable. Numerical results demonstrated that the proposed method can achieve the determination of the electrode distribution for electroosmotic micromixer, and the effectivity of the proposed method is confirmed by the postprocessing of the numerical results. In addition, this method can be extended to determine the electrode distribution for the electroosmotic micromixers with unsteady flow caused by the AC electroosmosis. And this will be investigated in the future.

\section*{Acknowledgements}
This work is supported by the National Natural Science Foundation of China (No. 51405465), the National High Technology Program of China (No. 2015AA042604), Science and Technology Development Plan of JiLin Province and Changchun City (No. 20140201011GX, 15SS12), and SKLAO Open Fund in CIOMP.

\end{document}